\documentclass[aps,prl,twocolumn]{revtex4-1}

\usepackage{amsmath}
\usepackage{graphicx,xcolor}
\usepackage{amsmath,amssymb,mathtools,slashed}
\usepackage{tikz-feynman}[luatex=true]

\usepackage{feynmp-auto}
\usepackage{simpler-wick}
\usepackage{kantlipsum}
\usepackage{float}
\usepackage{multirow}
\usepackage{rotating}

\usepackage[T1]{fontenc} 
\usepackage{etoolbox}
\usepackage{extarrows} 
\usepackage{graphicx}
\usepackage{dcolumn}
\usepackage{bm}
\usepackage{wasysym}
\usepackage{verbatim}

\usepackage{empheq}
\usepackage{multirow}
 
\usepackage{balance}

\usepackage{subfigure}
\usepackage[colorlinks=true,urlcolor=blue,linkcolor=blue]{hyperref}
\usepackage[capitalise]{cleveref}
\usepackage{siunitx}
\usepackage{enumerate}

\usepackage{array}
\usepackage{tcolorbox}
\usepackage{fontawesome}

\renewcommand{\vec}[1]{{\mathbf{#1}}}

\newcommand{\p}{\partial}

\newcommand{\bx}{{\bf{x}}}

\newcommand{\E}{\boldsymbol{e}}

\newcommand{\mH}{\mathcal{H}}

\newcommand{\an}{\quad \textmd{and} \quad }
\newcommand{\bs}{\boldsymbol{\sigma}}

\newcommand{\bea}{\begin{eqnarray}}
\newcommand{\eea}{\end{eqnarray}}

\newcommand{\bq}{{\bf{q}}}

\newcommand{\bk}{{\bf{k}}}

\usepackage{upgreek}

\newcommand{\mpl}{M_{\mbox{\tiny{Pl}}}}

\newcommand{\ev}[1]{\ensuremath{\left\langle #1 \right\rangle}}

\begin{document}

\title{Gravitational Wave-Induced Freeze-In of Fermionic Dark Matter}
\date{\today}
\author{Azadeh Maleknejad}
\email{azadeh.maleknejad@swansea.ac.uk}
\affiliation{Centre for Quantum Fields and Gravity, Swansea University,
Swansea SA2 8PP, United Kingdom}
\affiliation{Deutsches Elektronen-Synchrotron DESY,
Notkestra\ss e 85, 22607 Hamburg, Germany}
\affiliation{Institute of Theoretical Physics, Universit\"at Hamburg,
22761 Hamburg, Germany}

\author{Joachim Kopp}
\email{jkopp@cern.ch}
\affiliation{Theoretical Physics Department, CERN, 1211 Geneva 23, Switzerland}
\affiliation{PRISMA Cluster of Excellence \& Mainz Institute for Theoretical Physics, \\
             Johannes Gutenberg University, Staudingerweg 7, 55099 Mainz, Germany}

\preprint{KCL-PH-TH/2024-28, CERN-TH-2024-061, MITP-24-050}

\begin{abstract}
The minimal coupling of massless fermions to gravity does not allow for their gravitational production solely based on the expansion of the Universe. We argue that this changes in presence of realistic and potentially detectable stochastic gravitational wave backgrounds. We compute the resulting energy density of Weyl fermions at 1-loop using in--in formalism. If the initially massless fermions eventually acquire mass, this mechanism can explain the dark matter abundance in the Universe. Remarkably, it may be more efficient than conventional gravitational production of superheavy fermions.
\href{https://github.com/koppj/GW-freeze-in/}{\faGithub}
\end{abstract}

\maketitle

%
{\bf Introduction.}
The production of dark matter (DM), and possibly other feebly interacting particles, in the early Universe, remains a mystery. An intriguing possibility is that gravity itself serves as the mechanism responsible for the creation of DM in the Universe \cite{Ford:1986sy, Chung:1998zb, Parker_Toms_2009, Kolb:2023ydq}. However, the conventional mechanism requires very massive fields ($M \sim \SI{e14}{GeV}$) \cite{Kolb:2017jvz, Ema:2019yrd} and/or a a very hot plasma with temperatures $T_\text{reh}\gtrsim 10^{13}$ GeV \cite{Garny:2015sjg, Bernal:2018qlk, Clery:2021bwz}. Otherwise the fermions would enjoy at least approximate conformal symmetry, and their energy density would be a scaleless integral that vanishes \cite{Schwartz:2014sze}. Besides large masses, conformal symmetry can also be broken via interactions, for instance with SM fields or with the inflaton \cite{Greene:1998nh, Adshead:2018oaa, Maleknejad:2019hdr, Maleknejad:2020yys, Maleknejad:2020pec}. Alternatively, chiral gravitational waves (GWs) in inflation can produce Weyl fermions when parity is broken in Chern--Simons gravity \cite{Alexander:2004us} or by non-Abelian gauge fields in inflation \cite{Maleknejad:2016dci, Maleknejad:2014wsa, Adshead:2015jza, Caldwell:2017chz}.

The starting point of this letter is the observation that cosmic perturbations naturally and unavoidably break the conformal symmetry of Weyl fermions in General Relativity. This raises the question of whether such perturbations -- in particular in the form of stochastic gravitational waves -- can be responsible for the production of DM or other very weakly interacting particles in the early Universe. Similar questions were explored
in earlier works, e.g. \cite{Campos:1991ff}. Remarkably, as we will show here, the answer is Yes.

This is particularly relevant because the early Universe can be expected to be permeated by stochastic GWs. Numerous mechanisms for their production have been studied in detail, including gauge fields in inflation \cite{Sorbo:2011rz, Maleknejad:2016qjz, Komatsu:2022nvu}, first-order phase transitions \cite{Witten:1984rs, Schwaller:2015tja, Caprini:2015zlo, RoperPol:2023bqa}, primordial magnetic fields \cite{Brandenburg:2021aln, RoperPol:2021xnd}, preheating and gauge preheating \cite{Adshead:2019igv, Figueroa:2022iho}, cosmic strings \cite{Hindmarsh:1994re, Auclair:2019wcv}, etc. Stochastic GW backgrounds have been a hot topic of research for decades and their detection prospects have been thoroughly discussed. However, their role in DM freeze-in remains, to the best of our knowledge, a largely unexplored avenue. 

In this letter, we compute the energy density of Weyl fermions in the presence of a stochastic GW background in the early Universe at 1-loop using in--in formalism. Our key result will be that a GW background in the early Universe introduces new physical scales and naturally produces Weyl fermions. If these fermions later acquire a mass, they can play the role of the DM today. To introduce this new DM production mechanism and find analytical estimates, we consider a simple phenomenological broken power-law model for the GW spectrum, which provides a good fit to the results of simulations in many scenarios, e.g.\ phase transitions \cite{Caprini:2009yp} and primordial magnetic fields \cite{Caprini:2018mtu}. We expect that our result is generic, but accurately estimating the resulting fermion energy density for other sources of primordial GWs typically requires advanced modeling and simulations, which we leave for future work. A more detailed analysis can be found in the companion paper, Ref.~\cite{JHEP}.

%
{\bf Fermions in an Expanding Universe.}
The action of massless fermions in curved spacetime is
\begin{align}
    S_{\psi} =  \frac{i}{2} \int\!\text{d}^4x \, \sqrt{-g} \
               \big[ \boldsymbol{\uppsi}^\dag_L \bar{\bs}^\mu
                     \overset{\leftrightarrow}{\mathcal{D}}_\mu \boldsymbol{\uppsi}_L
               + \boldsymbol{\uppsi}^\dag_R \bs^\mu
                     \overset{\leftrightarrow}{\mathcal{D}}_\mu \boldsymbol{\uppsi}_R \big] ,
    \label{Eq:FullAction-F}
\end{align}
where $\boldsymbol{\uppsi}_{R,L}$ are the right-handed and left-handed Weyl fermion fields, respectively, $\mathcal{D}_\mu$ is the spinor covariant derivative, and $\overset{\leftrightarrow}{\mathcal{D}}_{\mu} \equiv \overset{\rightarrow}{\mathcal{D}}_{\mu} - \overset{\leftarrow}{\mathcal{D}}_{\mu}$.  In the Friedmann–-Lema\^itre–-Robertson–Walker (FLRW) metric of an expanding Universe, the effect of curvature can be absorbed by redefining the fermion field as $\boldsymbol{\Psi} \equiv a^{3/2} \boldsymbol{\uppsi}$, where $\boldsymbol{\Psi}$ is the canonically normalized field associated with $\boldsymbol{\uppsi}$ and $a$ is the time-dependent scale factor of the Universe. The fermion energy density is diluted like $a^{-4}$ with the expansion of the Universe as a consequence of the conformal symmetry of Weyl fermions. In the following, we will show that cosmic perturbations break this conformal symmetry and create Weyl fermions.

The metric of an expanding Universe permeated by a GW background is, up to first order in the GW amplitude,
\begin{align}
    ds^2 = -dt^2 + a^2(t) \big( \delta_{ij} + h_{ij} \big) dx_i dx_j,
    \label{eq:metric}
\end{align}
where $h_{ij}$ describes the GW, for which we choose the transverse--traceless gauge. We use the mostly-plus convention for the metric in this paper. The metric perturbation can be decomposed into Fourier modes as
\begin{align}
    h_{ij}(\tau,\vec{x}) &= \sum_{s=\pm} \int\!d^3q \, \mathrm{h}_{s,\bq}(\tau) \, \E^s_{ij}(\hat{\vec{q}})
                            \, e^{i \bq.\bx} \, \hat{a}^s_{\bq}  + c.c. \,,
    \label{eq:GW}
\end{align}
where the sum runs over the two circular polarization states of GWs,\footnote{Working with circular GW polarization states is more convenient here than using the plus and cross polarizations as a basis, even though the latter is more common in the GW literature.} and $\tau$ is the conformal time, related to physical time via $d\tau = dt / a(t)$. The Fourier coefficients $\mathrm{h}_{\pm,\bq}$ give the amplitude of each Fourier mode and polarization state, $\hat{a}^s_\bq$ denotes the canonically normalized graviton annihilation operators, and $\E^\pm_{ij}(\hat{\vec{q}})$ denotes the left-handed and right-handed circular polarization tensors of helicity $\pm 2$. In this work we consider unpolarized gravitational waves. 

\begin{figure}
    \centering
    \includegraphics[width=0.7\columnwidth]{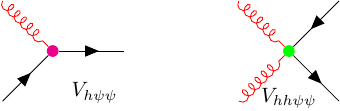}
    \caption{The graviton--fermion cubic and quartic vertices.}
    \label{fig:vertices}
\end{figure}

The first-order interaction Lagrangian between the Dirac field $\boldsymbol\Psi_{\!D} = (\boldsymbol\Psi_L, \boldsymbol\Psi_R)^\top$ and GWs is
\begin{align}
    \mathcal{L}^{(1)}_\text{int} = -\frac{i}{2a^4} h_{ij} \bar{\boldsymbol{\Psi}}_{\!D} \gamma^{i}
                                    \overset{\leftrightarrow}{\p}_j \boldsymbol{\Psi}_{\!D}.
    \label{Eq:L1-int}
\end{align}
It corresponds to the cubic vertex $V_{h\psi\psi}$ in \cref{fig:vertices}. At second order in the metric perturbation, the fermion--GW interaction is \footnote{Note that at second order in gravitational wave perturbations, the perturbed metric is $\delta g_{ij}=a^2(h_{ij} + \frac12 h_{ik} h_{jk})$.}
\begin{align}
    \mathcal{L}^{(2)}_\text{int} = -\frac{i}{16a^3}  {\bf{e}}^{\mu}_{~\alpha} h_{ij} \p_{\mu} h_{ik}
                                    \bar{\boldsymbol{\Psi}}_{\!D} \Gamma^{\alpha j k} \boldsymbol{\Psi}_{\!D},
    \label{Eq:L2-int}
\end{align}
which corresponds to the quartic vertex $V_{hh\psi\psi}$ in \cref{fig:vertices}. Here $\Gamma^{\alpha j k}$ is the totally antisymmetrized product of three gamma matrices, and ${\bf{e}}^\mu_{~\alpha}$ are the tetrads.  Note that we have written \cref{Eq:L1-int,Eq:L2-int} for Dirac fields; it is straightforward from these expressions to find the associated interaction vertices for Weyl fermions of either chirality. For the derivation of \cref{Eq:L1-int} and \cref{Eq:L2-int} see Appendix~B of \cite{JHEP}.

In the following, we focus on right-handed Weyl fermions; left-handed Weyl fermions be can treated completely analogously, and Dirac (Majorana) fermions can be easily expressed in terms of two (one) Weyl fermions. For simplicity, we omit the subscript $R$. The Weyl fermion field can be decomposed as
\begin{align}
    \boldsymbol{\Psi} =
        \int\! d^3k \Big[
            {\bf{U}}_{\bf{k}}(\tau) \, \hat{b}_{\bk} e^{i{\bf{k}}.{\bf{x}}}
         +  {\bf{V}}_{\bf{k}}(\tau) \, \hat{c}^{\dagger}_{\bk}
                e^{-i{\bf{k}}.{\bf{x}}} \Big] \,,
    \label{eq:Psi-Weyl}
\end{align}
where $\hat{b}_{\bk}$ denotes the particle annihilation operators and $\hat{c}^{\dagger}_{\bk}$ is the antiparticle creation operators. The corresponding spinors are ${\bf{U}}_{\bf{k}}(\tau)$ and ${\bf{V}}_{\bf{k}}(\tau)$; they are related by CP symmetry. At zeroth-order in $h_{ij}$ we have\footnote{Given that helicity and chirality are equivalent for free fermions, the zeroth-order solution for left-handed Weyl fermions is similar, but with ${\bf{E}}^+_{\bk}$ replaced by ${\bf{E}}^-_{\bk}$.}
\begin{align}
    {\bf{U}}^{(0)}_{{\bf{k}}} = \frac{e^{-ik\tau}}{\sqrt{2}(2\pi)^{\frac32}} {\bf{E}}^+_{\bk}
                                      \an  
    {\bf{V}}^{(0)}_{{\bf{k}}} = \frac{e^{ik\tau}}{\sqrt{2}(2\pi)^{\frac32}} {\bf{E}}^-_{-\bk},
    \label{eq:U0-V0}
\end{align}
where $k \equiv |\bk|$ and ${\bf{E}}^{\pm}_{\bf{k}}$ are the helicity eigenstates \cite{JHEP}. At zeroth order in the gravitational perturbation, the field equation of  Weyl fermions is not time-dependent, so no particle production occurs \cite{Kolb:2023ydq}.

As shown in the companion paper, Ref.~\cite{JHEP}, the contribution of the quartic interaction $V_{hh\psi\psi}$ to the fermion energy density vanishes for unpolarized GWs, therefore we will in the following neglect the quartic vertex.

We remark that if the GW background is chiral, the triangle diagram corresponding to the (global) gravitational anomaly should also be considered \cite{Alvarez-Gaume:1983ihn}, see for instance \cite{Alexander:2004us, Maleknejad:2016dci, Maleknejad:2024vvf}.

%
{\bf Fermion Production by Stochastic Gravitational Waves at 1-Loop.}
Consider the DM energy density
\begin{align}
     \rho_{\psi} = \frac{i}{2a^4} \boldsymbol{\Psi}^{\dag} \overset{\leftrightarrow}{\p}_{\tau}
                      \boldsymbol{\Psi}.
    \label{eq:Tmunu}
\end{align}
To determine this quantity in a GW background, we evaluate the expectation value of $\rho_{\psi}$ in the in--in formalism. Having established above that the contribution of $V_{hh\psi\psi}$ vanishes for unpolarized GWs, we evaluate the contribution of $ \mathcal{L}^{(1)}_\text{int}$  ($V_{h\psi\psi}$ in \cref{fig:vertices}), i.e., 
\begin{align}
    \langle\rho_{\psi}\rangle  \equiv \Big\langle  \int^\tau_{\tau_\text{in}} \! dt'' 
                      \, H_\text{int}(\tau'')  \,
                      \rho_{\psi,I}(\tau) \,
                      \int^\tau_{\tau_\text{in}} \! dt' \, H_\text{int}(\tau')  \Big\rangle,
    \label{Eq:rho-1}
    \end{align}
where $\tau_{\text{in}}$ is the initial time. Here, the operator $\rho_{\psi,I}(\tau)$ is $\rho_{\psi}(\tau,\bx)$ in the interaction picture, and the interaction Hamiltonian is
\begin{align}
    H_\text{int}(t) = -\int \! d^3x \, a^3(t) \, \mathcal{L}^{(1)}_\text{int}(t,\bx).
\end{align}
%
Evaluating \cref{Eq:rho-1} involves Wick-contracting of the fermion and graviton fields appearing in $H_\text{int}$ and $\rho_{\psi,I}$. Diagrammatically, these contractions can be represented as a loop diagram. Note that we do not evaluate the graviton two-point function $\ev{ \mathrm{h}_{s',\bq'}(\tau'') \mathrm{h}^{*}_{s,\bq}(\tau') }$ as a propagator of a quantum field, but as an expectation value of a classical field.

We present here a concise overview of the steps taken, deferring the detailed computation to the companion paper \cite{JHEP}.  Assuming an unpolarized GW background and imposing its statistical isotropy, we have
\begin{align}
    \ev{ \mathrm{h}_{s',\bq'}(\tau'') \mathrm{h}^{*}_{s,\bq}(\tau') }
        = \delta^3(\bq'-\bq) \, \delta_{ss'} \ev{ \mathrm{h}_{q}(\tau'') \, \mathrm{h}^{*}_{q} (\tau')}.
\end{align}
With this, one can show that the fermion energy density is \cite{JHEP} \footnote{We have already performed the quantum expectation value over the fermionic fields, and the remaining averaging on GW is purely stochastic.} (here $\omega = \sqrt{k^2+q^2-2kq\cos\theta}$)
\begin{align}
    \langle \rho_{\psi} \rangle &= \frac{1/4\pi }{a^4(\tau)}
        \int\!d\tau' \int\!d\tau''
        \int_0^{\infty} \! q^2 dq   \int_0^q \! k^4 dk \, \big\langle \mathrm{h}_{q}(\tau'') \mathrm{h}^*_{q}(\tau') \big\rangle \nonumber\\
    & \times  \int \! d\theta \, \sin^3\theta \, e^{i(k+\omega)(\tau'-\tau'')} (k+\omega)       \nonumber\\
    &\times  \Big[ 2 - \frac{\sin^2\theta((\omega+k)^2+q^2)}{2\omega(\omega+k-q\cos\theta)}\Big]  + c.c. \,.
    \label{Eq:rho-tot}
\end{align} 
The integral in $k$, which in principle extends from 0 to $\infty$ is cut off at scales around $q$, implying that fermions of momentum $k$ can only be created by GW modes with $q \gtrsim k$. The reason is that a long-wavelength GW mode with $q \ll k$ is seen by the fermion field as spatially quasi-homogeneous and can therefore be absorbed into a (large) gauge transformation \footnote{A long-wavelength GW mode (homogeneous tensor mode) $h^\text{long}_{ij}(t,\bx) \approx C_{ij}(t)$, can be recast as a change in coordinates $h^{\text{long}}_{ij} \mapsto 0$ and $x^i \mapsto x^{i'} = e^{h^{\text{long}}_{ij}} x^j$ \cite{Weinberg:2008zzc,Senatore:2012nq}. In other words, it can be considered as a large gauge transformation. As a result, after this coordinate transformation, a short-wavelength fermion mode in the presence of long-wavelength GWs becomes a free fermion, i.e., \smash{$\Psi_\text{short}(t,\bx) \big\rvert_{h_{ij}^{\text{long}}} = \Psi'(t,\bx') \big\rvert_0$}, which implies $\langle \rho_{\psi,\text{short}}(x) \rangle \big\rvert_{h_{ij}^\text{long}} = \langle \rho_{\psi,\text{short}}(x') \rangle \big\rvert_0 = 0$.}.
Here, as an approximation, we introduce the cutoff at $k=q$ by hand.

In analogy to $\langle \rho_{\psi} \rangle$, one can also calculate the pressure density of the dark fermions, which is found to be
\begin{align}
    \langle P_{\psi} \rangle = \frac13 \langle \rho_{\psi}\rangle 
                             \propto \frac{1}{a^4(\tau)}.
    \label{eq:P}
\end{align}
This shows that the fermions produced via gravitational wave-induced freeze-in indeed behave like radiation.

%
{\bf Stochastic Gravitational Wave Backgrounds.}
To be able to evaluate the remaining integrals in \cref{Eq:rho-tot}, we need to model the stochastic GW background. This background can originate from a variety of different processes, among the most important of which are: i) cosmological first-order phase transitions. Such transitions proceed through the nucleation and subsequent expansion of bubbles of the new phase inside the old phase. They generate GWs in bubble collisions and through sound waves and turbulence in the cosmic fluid \cite{Witten:1984rs, Schwaller:2015tja, Caprini:2015zlo, RoperPol:2023bqa}; ii) primordial magnetic fields \cite{Brandenburg:2021aln, RoperPol:2021xnd}; iii) preheating and gauge preheating \cite{Adshead:2019igv, Figueroa:2022iho}, and iv) cosmic strings \cite{Hindmarsh:1994re, Auclair:2019wcv}.  The GW power spectrum at a given (conformal) time $\tau$ depends on the dynamics of the GW production process and on the expansion history of the Universe. We separate these two dependencies and write
\begin{align}
    \mathrm{h}_{s,\bq}(\tau) = a^{-1}(\tau) \, \mathcal{T}(\tau,q) \, \mathrm{h}_{s,\bq,0},
    \label{eq:hsq-parameterization}
\end{align}
where $\mathrm{h}_{s\bq,0}$ is the GW spectral amplitude today at $\tau=\tau_0$, and $\mathcal{T}(\tau,q)$ is a transfer function that describes the dynamics of the GW production process and the gradual build-up of the GW background. We parameterize the transfer function for $\tau>\tau_\text{in}$ as
\begin{align}
    \mathcal{T}(\tau,q) \approx \big( 1 - e^{-\pi \beta(\tau-\tau_\text{in})} \big) \, e^{-iq\tau},
    \label{eq:transfer-function}
\end{align}
where $\beta^{-1}$ is the characteristic time scale associated with the process that sources the GW background. We denote the conformal times associated with the start and end of GW production as $\tau_\text{in}$ and $\tau_*$, respectively. We assume that $\beta^{-1}$ is shorter than a Hubble time, i.e., $\beta/\mathcal{H}_* > 1$, where $\mathcal{H}_* \equiv a(\tau_*) \, H(\tau_*)$, and $H(\tau_*)$ is the value of the Hubble parameter at conformal time $\tau_*$.

Describing the dynamics of GW production through any mechanism and predicting the final spectrum $\mathrm{h}_{s\bq,0}$ typically requires sophisticated numerical simulations. Nevertheless, the results can typically be approximated by relatively simple analytical fitting functions. Here, we will use a broken power law parameterization. Expressed in terms of the fractional cosmological energy density of GWs today, $\Omega_{\text{gw},0}(q) = 2\pi q^5 / (3 H_0^2) \sum_s \langle \mathrm{h}_{s,\bq}(\tau_0) \, \mathrm{h}_{s,\bq}^*(\tau_0) \rangle$, our ansatz reads \footnote{The energy density of GWs is defined as $\rho_\text{gw}  = \frac{1}{32\pi G} \big\langle \dot{h}_{ij}\, \dot{h}^{ij}\big\rangle$ and its spectral energy density is $\Omega_\text{gw} = \frac{1}{\mpl^2} \frac{d\rho_\text{gw}}{3H^2 d\ln q} $.}
\begin{align}
    \Omega_{\text{gw},0}(q) \approx
        \begin{cases}
            \Omega_\text{peak} \, \big( \frac{q}{q_\text{peak}} \big)^m
                &  q_\text{min} < q < q_\text{peak} , \\
            \Omega_\text{peak} \, \big( \frac{q}{q_\text{peak}} \big)^{-n}
                &  q_\text{peak} < q < q_\text{max} ,
                \end{cases}
    \label{eq:Omega-PT}
\end{align}
with $m, n > 0$. This ansatz corresponds to a peaked spectrum with a maximum spectral energy density $\Omega_\text{peak}$ at momentum (frequency) $q_\text{peak}$. There are three relevant physical scales here: at low frequencies, the horizon scale $q_\text{min} = \mathcal{H}_*$ provides a natural cutoff; at high frequencies, the spectrum extends up to the smallest physical scales associated with the source, typically of order $q_\text{max}\approx a_* T_*$ (where $T_*$ is the temperature of the hot early Universe plasma at $\tau_*$); finally, $q_\text{peak}$ is a typical scale characterizing the source. \Cref{eq:Omega-PT} provides a good fit to the results of simulations in many scenarios, e.g.\ phase transitions  \cite{Caprini:2009yp, Durrer:2010xc} and primordial magnetic fields \cite{Caprini:2018mtu, RoperPol:2022iel}. For GW from phase transitions, the spectral index at frequencies below the peak is $m \approx 3$ (imposed by causality), while the one at high frequencies varies across the range $n \sim (1-4)$, depending on whether GW emission is dominated by bubble collisions, sound waves, or turbulence. Which of these contributions is most relevant depends on the details of the phase transition (degree of temporal coherence, runaway bubbles vs.\ non-runaway bubbles \cite{Caprini:2009yp, Durrer:2010xc, Breitbach:2018ddu, Breitbach:2018kma}).

\Cref{eq:Omega-PT} alone is not sufficient to evaluate the final fermion energy density which, according to \cref{Eq:rho-tot} depends not simply on the GW power spectrum, but on their \emph{unequal-time} two-point correlation function, $\ev{\mathrm{h}_{s,\bq}(\tau'') \mathrm{h}^*_{s,\bq}(\tau')}$, which is considerably more challenging to determine than $\Omega_{\text{gw},0}(q)$. However, it is again possible to use simple phenomenological models. We write \cite{Scully_Zubairy_1997}
\begin{align}
    \ev{\mathrm{h}_{s,\bq}(\tau'') \, \mathrm{h}^*_{s,\bq}(\tau')}
      =  \gamma_q(\Delta\tau) \,
         \sqrt{\ev{\lvert \mathrm{h}_{s,\bq}(\tau')\rvert^2}
               \ev{\lvert \mathrm{h}_{s,\bq}(\tau'')\rvert^2}} \,,
    \label{eq:coherent}
\end{align}
where $\Delta\tau = \lvert \tau' - \tau''\rvert$ and $\gamma_q(\Delta\tau)$ parameterizes the degree of temporal coherence. We distinguish between two extreme cases, namely fully incoherent GWs, for which $\gamma_q(\Delta\tau) = \Delta\eta \, \delta(\tau'-\tau'')$, and the characteristic coherence time $\Delta\eta$ is much shorter than the dynamical time scales in the system, i.e. $\Delta\eta \ll \beta ^{-1}$ \cite{Caprini:2009yp}; and fully coherent GWs, for which $\gamma_q( \Delta\tau) = 1$. As discussed in Refs.~\cite{Caprini:2009fx, Caprini:2009yp}, bubble collisions during a first-order phase transition can be considered fully coherent, that is, deterministic in time. In contrast, turbulence and magnetic field are partially coherent GW sources.

%
{\bf Dark Matter Relic Density.} We are now ready to calculate the fermion energy density at a time $\tau > \tau_*$. To do so, we plug the parameterization of the GW background from \cref{eq:hsq-parameterization,eq:transfer-function,eq:Omega-PT,eq:coherent} into \cref{Eq:rho-tot}. 
We find (see Ref.~\cite{JHEP}, for details) 
\begin{align}
    \rho_{\psi}(\tau) \approx \mathcal{C} \, \Big( \frac{q_\text{peak}}{a(\tau)} \Big)^4 
                                      \bigg( \frac{H_0}{\mathcal{H}_*} \bigg)^2
                                      z_*^2 \, \Omega_\text{peak},
 \label{eq:rho-psi-}                                     
\end{align}
where $\mathcal{C}$  is a dimensionless parameter differing for coherent and incoherent GWs as
\begin{align}
    \mathcal{C}_\text{incoh} &\approx 0.10 \Delta\eta \, \beta
        \begin{cases}
            \frac{1}{4-n} \big(\frac{q_\text{max}}{q_\text{peak}} \big)^{4-n}
                    & \text{for $n<4$},\\
            \ln\Big( \frac{q_\text{max}}{q_\text{peak}} \Big)
                    & \text{for $n=4$}, \\
            \frac{(n+m)}{(4+m)(n-4)}
                    & \text{for $n>4$}
        \end{cases}.
    \label{Eq:Cmn-incoh}
\intertext{for incoherent GWs, and}
    \mathcal{C}_{\text{coh}} &\approx
            0.0084 + 0.023
            \begin{cases}
                \frac{1}{n-2} \big[ 1\! -\! \big(\frac{q_\text{max}}{q_\text{peak}}\big)^{2-n} \big]
                                         & \text{\!\!for $n \neq 2$}\\
                \ln\frac{q_{\text{max}}}{q_\text{peak}}
                                         & \text{\!\!for $n = 2$}
            \end{cases}
    \label{Eq:Cmn-coh}
\end{align}
for coherent sources. To keep the second of these expressions short, we have taken $m=3$ and we have set $\beta = q_\text{peak}$. The full expression can be found in Ref.~\cite{JHEP}. To provide insight into the qualitative form of \cref{eq:rho-psi-}, we note that it scales as $a^{-4}$, characteristic of massless fields, with $q_\text{peak}^4$ dependence arising from dimensional analysis, and the GW energy density at production, $\Omega_\text{peak}^* = \frac{\rho_{\text{gw},*}}{3H_*^2\mpl^2} = z_*^2\big(\frac{H_0}{\mathcal{H}_*}\big)^2\Omega_\text{peak}$.

For both coherent and incoherent GW sources, we find that $\mathcal{C}$ depends strongly on the slope of the spectrum at high frequencies. $\mathcal{C}_\text{incoh}$ is $\mathcal{O}(1)$ for $n \geq 4$, but depends strongly on $q_\text{max} / q_\text{peak}$ for $n < 4$. On the other hand, $\mathcal{C}_\text{coh}$ is $\mathcal{O}(1)$ for $n \geq 2$, but has the strong dependence on $q_\text{peak}/q_\text{max}$ for $n<2$. That is due to the shape of the spectral number density of fermions. It exhibits a broken power-law: $n_k\propto (k/q_\text{peak})^4$ for $k<q_\text{peak}$ and $n_k\propto (k/q_\text{peak})^{\alpha-n}$ for $q_\text{peak}<k<q_\text{max}$ with $\alpha_\text{coh}=1$  and $\alpha_\text{incoh}=3$.  The break at $k=q_\text{peak}$ marks the transition between these two regimes.

Note that in our calculations we have neglected the backreaction of fermion production onto the GW background. This is justified by noting that $\rho_{\psi} / \rho_\text{gw} \sim \mathcal{C} q_\text{peak}^4 / (\mathcal{H}_*^2 \mpl^2) \ll 1$.

To derive the fractional energy density of dark fermions today, $\Omega_{\psi,0}$ we assume that the fermions, while effectively massless at the time of production (where $T \simeq T_*$), have non-negligible mass $M$ today, leading to $\frac{\rho_{\psi}(\tau)}{\rho_\text{gw}(\tau)}\propto a(\tau)$.  The initially negligible fermion to GW energy density increases with time and eventually can be very large. Fermions can either remain massless during the phase transition, acquiring mass through a Higgs mechanism at lower temperatures, or have a mass from the start, but $M \lesssim H_*$.  The fermion energy density today is \footnote{Note that $\Omega_\text{peak}$ may be related to $T_*$ and the GW background properties in a model-dependent manner. This work focuses on a broadly applicable framework, leaving a detailed exploration to future numerical studies.}
\begin{align}
    \Omega_{\psi,0}
       & \equiv \frac{\rho_{\psi}(\tau_0)}{3 \mpl^2 H_0^2}
       \approx  0.36 \, \mathcal{C} \,
                \bigg( \frac{g_*}{106.75}\bigg)^{\frac43}
                \bigg( \frac{\Omega_\text{peak}}{10^{-6}} \bigg)  \nonumber\\
       &\times  \bigg( \frac{M}{T_*} \bigg) \bigg(\frac{q_\text{peak}/ \mH_*}{100} \bigg)^4
                \bigg( \frac{T_*}{\SI{3e11}{GeV}} \bigg)^5,
    \label{eq:OmegaD}
\end{align}
where $\mpl= 2.4 \times 10^{18}$ GeV is the reduced Planck mass. The most important dependencies of $\Omega_{\psi,0}$ are on the GW peak frequency $q_\text{peak}$ (due to the strong momentum dependence of $\ev{\rho_\psi}$, see \cref{Eq:rho-tot}, which ultimately stems from the derivative in the interaction vertex, \cref{Eq:L1-int}) and on the temperature $T_*$ (due to the redshift of the GW spectrum). Typical values for the ratio between the GW frequency and the Hubble scale (or, equivalently the Hubble radius and the GW wavelength) are $q_\text{peak} / \mathcal{H}_* \gtrsim \num{e2}$ for first-order phase transitions \cite{Caprini:2009yp}. For the number of relativistic degrees of freedom at $T_*$, we have used the SM value $g_* = 106.75$.

\begin{figure}
    \centering
    \includegraphics[width=\columnwidth]{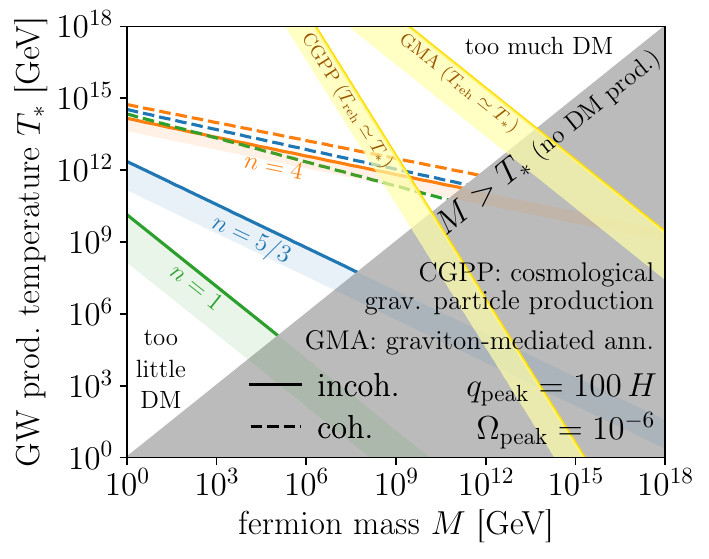}
    \caption{GW-induced freeze-in of dark matter for a GW background with a broken power-law spectrum. Colored lines show the phase transition temperature $T_*$ and DM mass $M$ required to explain the observed DM density. We have assumed a spectral index $m=3$ below the peak frequency, $q_\text{peak}$, while above the peak we have used $n=1$ (green, e.g.\ from bubble collisions in a first-order phase transition), $n=5/3$ (blue, e.g.\ from turbulence), and $n=4$ (orange, e.g.\ from sound waves) as benchmark values. Below the lines, GW-induced freeze-in can still contribute a fraction of the DM. The bottom edges of the shaded bands indicate where that fraction is 1\%. For comparison, we show in yellow the parameter region in which conventional cosmological production of supermassive fermions by the expansion of the Universe (``CGPP'') \cite{Kolb:2017jvz, Ema:2019yrd, Kolb:2023ydq} or graviton-mediated annihilation (``GMA'') \cite{Bernal:2018qlk, Clery:2021bwz} yield the correct relic density. Inside the gray area, fermions are massive already at the time $T_*$ of GW production, so our mechanism is not applicable. \emph{Gravitational-wave induced freeze-in can successfully explain the observed DM relic abundance in large swaths of parameter space, favoring $T_*$ well above the electroweak scale.}}
    \label{fig:OmegaD}
\end{figure}

We plot our result for $\Omega_{\psi,0}$ in \cref{fig:OmegaD}. We see that our mechanism can explain the observed DM density in the Universe for a wide range of DM masses and temperature scales, typically favoring $T_*$ well above the electroweak scale, but safely below the Planck scale, and well below the scales required for cosmological gravitational production of supermassive fermions \cite{Kolb:2017jvz, Kolb:2023ydq} and graviton-mediated annihalations in a very high temperature plasma \cite{Bernal:2018qlk, Clery:2021bwz}  (yellow bands in \cref{fig:OmegaD}, see \cite{JHEP} for details). In producing the plot, we have chosen particular values of $\Omega_\text{peak}$ and $q_\text{peak}$ (with $\Omega_\text{peak}$ below the Big Bang Nucleosynthesis bound, $\Omega_\text{peak} \lesssim \num{3.5e-6} \, m \, n / (m + n)$) \cite{Maggiore:2018sht}, but the scaling at different $\Omega_\text{peak}$, $q_\text{peak}$ can be read off immediately from \cref{eq:OmegaD}.

Note that the values of $T_*$ and $q_\text{peak}$ for which our model explains the DM relic density correspond to GW peak frequencies today, $f_\text{peak}$, of order kHz to GHz. The lower end of this range falls within the sensitivity range of future interferometric gravitational wave detectors like Einstein Telescope \cite{Maggiore:2019uih} and Cosmic Explorer \cite{Evans:2021gyd}. Signals at higher frequencies are currently still out of reach, though many novel detection concepts for GWs above the kHz band have been proposed and are seeing rapid progress \cite{Aggarwal:2020olq}.

{\bf Summary and Outlook.}
In this letter, we have reported the surprising connection between stochastic GW backgrounds and the gravitational production of fermions in the early Universe. Notably, while the expansion of the Universe alone cannot change the number density of massless fermions, in the presence of cosmic perturbations like a classical background of GWs it can. The crucial point here is that a GW background introduces new scales in the system and breaks the fermions' conformal invariance. Our mechanism is also valid in asymptotically flat spacetimes, though the resulting fermion abundance in this case is cosmologically insignificant. The reason is that, in flat spacetime, GW sources are typically local, and the GW strain decays as $r^{-1}$ away from the source \cite{Maggiore:2007ulw, Strominger:2017zoo}. 

As an important application of GW-induced freeze-in, we have argued that, if the initially massless (or effectively massless) fermions later acquire mass, they can constitute all or part of the dark matter in the Universe. In addition, GW-induced freeze-in may constitute an important pathway for producing other extremely weakly interacting fermion species such as right-handed neutrinos.

The next step in advancing this direction of research will be to go beyond our analytical estimates and to use numerical simulations to refine the precision of our predictions. Such analyses could, for example, be applied to gravitational waves sourced by primordial magnetic fields as well as to fluctuations generated during inflation.


\begin{acknowledgments}
{\bf{Acknowledgments:} } We are grateful to Nima Arkani-Hamed, 
Samuel Abreu, Cliff Burgess, Chiara Caprini, Timothy Cohen, Raphael Flauger, Danial Green, Edward Kolb, Eiichiro Komatsu, Andrew Long, Juan Maldacena, Marco Simonovich, and Alexander Zhiboedov, for useful discussions. A.M. would like to thank Hengameh Bagherian and Alberto Roper Pol for their collaboration on related projects. The work of A.M. is supported by the Royal Society University Research Fellowship, Grant No. RE22432. 
\end{acknowledgments}


\bibliographystyle{apsrev4-1}
\bibliography{ref.bib}

\begin{thebibliography}{60}%
\makeatletter
\providecommand \@ifxundefined [1]{%
 \@ifx{#1\undefined}
}%
\providecommand \@ifnum [1]{%
 \ifnum #1\expandafter \@firstoftwo
 \else \expandafter \@secondoftwo
 \fi
}%
\providecommand \@ifx [1]{%
 \ifx #1\expandafter \@firstoftwo
 \else \expandafter \@secondoftwo
 \fi
}%
\providecommand \natexlab [1]{#1}%
\providecommand \enquote  [1]{``#1''}%
\providecommand \bibnamefont  [1]{#1}%
\providecommand \bibfnamefont [1]{#1}%
\providecommand \citenamefont [1]{#1}%
\providecommand \href@noop [0]{\@secondoftwo}%
\providecommand \href [0]{\begingroup \@sanitize@url \@href}%
\providecommand \@href[1]{\@@startlink{#1}\@@href}%
\providecommand \@@href[1]{\endgroup#1\@@endlink}%
\providecommand \@sanitize@url [0]{\catcode `\\12\catcode `\$12\catcode `\&12\catcode `\#12\catcode `\^12\catcode `\_12\catcode `\%12\relax}%
\providecommand \@@startlink[1]{}%
\providecommand \@@endlink[0]{}%
\providecommand \url  [0]{\begingroup\@sanitize@url \@url }%
\providecommand \@url [1]{\endgroup\@href {#1}{\urlprefix }}%
\providecommand \urlprefix  [0]{URL }%
\providecommand \Eprint [0]{\href }%
\providecommand \doibase [0]{http://dx.doi.org/}%
\providecommand \selectlanguage [0]{\@gobble}%
\providecommand \bibinfo  [0]{\@secondoftwo}%
\providecommand \bibfield  [0]{\@secondoftwo}%
\providecommand \translation [1]{[#1]}%
\providecommand \BibitemOpen [0]{}%
\providecommand \bibitemStop [0]{}%
\providecommand \bibitemNoStop [0]{.\EOS\space}%
\providecommand \EOS [0]{\spacefactor3000\relax}%
\providecommand \BibitemShut  [1]{\csname bibitem#1\endcsname}%
\let\auto@bib@innerbib\@empty
\bibitem [{\citenamefont {Ford}(1987)}]{Ford:1986sy}%
  \BibitemOpen
  \bibfield  {author} {\bibinfo {author} {\bibfnamefont {L.~H.}\ \bibnamefont {Ford}},\ }\href {\doibase 10.1103/PhysRevD.35.2955} {\bibfield  {journal} {\bibinfo  {journal} {Phys. Rev. D}\ }\textbf {\bibinfo {volume} {35}},\ \bibinfo {pages} {2955} (\bibinfo {year} {1987})}\BibitemShut {NoStop}%
\bibitem [{\citenamefont {Chung}\ \emph {et~al.}(1998)\citenamefont {Chung}, \citenamefont {Kolb},\ and\ \citenamefont {Riotto}}]{Chung:1998zb}%
  \BibitemOpen
  \bibfield  {author} {\bibinfo {author} {\bibfnamefont {D.~J.~H.}\ \bibnamefont {Chung}}, \bibinfo {author} {\bibfnamefont {E.~W.}\ \bibnamefont {Kolb}}, \ and\ \bibinfo {author} {\bibfnamefont {A.}~\bibnamefont {Riotto}},\ }\href {\doibase 10.1103/PhysRevD.59.023501} {\bibfield  {journal} {\bibinfo  {journal} {Phys. Rev. D}\ }\textbf {\bibinfo {volume} {59}},\ \bibinfo {pages} {023501} (\bibinfo {year} {1998})},\ \Eprint {http://arxiv.org/abs/hep-ph/9802238} {arXiv:hep-ph/9802238} \BibitemShut {NoStop}%
\bibitem [{\citenamefont {Parker}\ and\ \citenamefont {Toms}(2009)}]{Parker_Toms_2009}%
  \BibitemOpen
  \bibfield  {author} {\bibinfo {author} {\bibfnamefont {L.}~\bibnamefont {Parker}}\ and\ \bibinfo {author} {\bibfnamefont {D.}~\bibnamefont {Toms}},\ }\href@noop {} {\emph {\bibinfo {title} {Quantum Field Theory in Curved Spacetime: Quantized Fields and Gravity}}},\ Cambridge Monographs on Mathematical Physics\ (\bibinfo  {publisher} {Cambridge University Press},\ \bibinfo {year} {2009})\BibitemShut {NoStop}%
\bibitem [{\citenamefont {Kolb}\ and\ \citenamefont {Long}(2023)}]{Kolb:2023ydq}%
  \BibitemOpen
  \bibfield  {author} {\bibinfo {author} {\bibfnamefont {E.~W.}\ \bibnamefont {Kolb}}\ and\ \bibinfo {author} {\bibfnamefont {A.~J.}\ \bibnamefont {Long}},\ }\href@noop {} {\  (\bibinfo {year} {2023})},\ \Eprint {http://arxiv.org/abs/2312.09042} {arXiv:2312.09042 [astro-ph.CO]} \BibitemShut {NoStop}%
\bibitem [{\citenamefont {Kolb}\ and\ \citenamefont {Long}(2017)}]{Kolb:2017jvz}%
  \BibitemOpen
  \bibfield  {author} {\bibinfo {author} {\bibfnamefont {E.~W.}\ \bibnamefont {Kolb}}\ and\ \bibinfo {author} {\bibfnamefont {A.~J.}\ \bibnamefont {Long}},\ }\href {\doibase 10.1103/PhysRevD.96.103540} {\bibfield  {journal} {\bibinfo  {journal} {Phys. Rev. D}\ }\textbf {\bibinfo {volume} {96}},\ \bibinfo {pages} {103540} (\bibinfo {year} {2017})},\ \Eprint {http://arxiv.org/abs/1708.04293} {arXiv:1708.04293 [astro-ph.CO]} \BibitemShut {NoStop}%
\bibitem [{\citenamefont {Ema}\ \emph {et~al.}(2019)\citenamefont {Ema}, \citenamefont {Nakayama},\ and\ \citenamefont {Tang}}]{Ema:2019yrd}%
  \BibitemOpen
  \bibfield  {author} {\bibinfo {author} {\bibfnamefont {Y.}~\bibnamefont {Ema}}, \bibinfo {author} {\bibfnamefont {K.}~\bibnamefont {Nakayama}}, \ and\ \bibinfo {author} {\bibfnamefont {Y.}~\bibnamefont {Tang}},\ }\href {\doibase 10.1007/JHEP07(2019)060} {\bibfield  {journal} {\bibinfo  {journal} {JHEP}\ }\textbf {\bibinfo {volume} {07}},\ \bibinfo {pages} {060} (\bibinfo {year} {2019})},\ \Eprint {http://arxiv.org/abs/1903.10973} {arXiv:1903.10973 [hep-ph]} \BibitemShut {NoStop}%
\bibitem [{\citenamefont {Garny}\ \emph {et~al.}(2016)\citenamefont {Garny}, \citenamefont {Sandora},\ and\ \citenamefont {Sloth}}]{Garny:2015sjg}%
  \BibitemOpen
  \bibfield  {author} {\bibinfo {author} {\bibfnamefont {M.}~\bibnamefont {Garny}}, \bibinfo {author} {\bibfnamefont {M.}~\bibnamefont {Sandora}}, \ and\ \bibinfo {author} {\bibfnamefont {M.~S.}\ \bibnamefont {Sloth}},\ }\href {\doibase 10.1103/PhysRevLett.116.101302} {\bibfield  {journal} {\bibinfo  {journal} {Phys. Rev. Lett.}\ }\textbf {\bibinfo {volume} {116}},\ \bibinfo {pages} {101302} (\bibinfo {year} {2016})},\ \Eprint {http://arxiv.org/abs/1511.03278} {arXiv:1511.03278 [hep-ph]} \BibitemShut {NoStop}%
\bibitem [{\citenamefont {Bernal}\ \emph {et~al.}(2018)\citenamefont {Bernal}, \citenamefont {Dutra}, \citenamefont {Mambrini}, \citenamefont {Olive}, \citenamefont {Peloso},\ and\ \citenamefont {Pierre}}]{Bernal:2018qlk}%
  \BibitemOpen
  \bibfield  {author} {\bibinfo {author} {\bibfnamefont {N.}~\bibnamefont {Bernal}}, \bibinfo {author} {\bibfnamefont {M.}~\bibnamefont {Dutra}}, \bibinfo {author} {\bibfnamefont {Y.}~\bibnamefont {Mambrini}}, \bibinfo {author} {\bibfnamefont {K.}~\bibnamefont {Olive}}, \bibinfo {author} {\bibfnamefont {M.}~\bibnamefont {Peloso}}, \ and\ \bibinfo {author} {\bibfnamefont {M.}~\bibnamefont {Pierre}},\ }\href {\doibase 10.1103/PhysRevD.97.115020} {\bibfield  {journal} {\bibinfo  {journal} {Phys. Rev. D}\ }\textbf {\bibinfo {volume} {97}},\ \bibinfo {pages} {115020} (\bibinfo {year} {2018})},\ \Eprint {http://arxiv.org/abs/1803.01866} {arXiv:1803.01866 [hep-ph]} \BibitemShut {NoStop}%
\bibitem [{\citenamefont {Clery}\ \emph {et~al.}(2022)\citenamefont {Clery}, \citenamefont {Mambrini}, \citenamefont {Olive},\ and\ \citenamefont {Verner}}]{Clery:2021bwz}%
  \BibitemOpen
  \bibfield  {author} {\bibinfo {author} {\bibfnamefont {S.}~\bibnamefont {Clery}}, \bibinfo {author} {\bibfnamefont {Y.}~\bibnamefont {Mambrini}}, \bibinfo {author} {\bibfnamefont {K.~A.}\ \bibnamefont {Olive}}, \ and\ \bibinfo {author} {\bibfnamefont {S.}~\bibnamefont {Verner}},\ }\href {\doibase 10.1103/PhysRevD.105.075005} {\bibfield  {journal} {\bibinfo  {journal} {Phys. Rev. D}\ }\textbf {\bibinfo {volume} {105}},\ \bibinfo {pages} {075005} (\bibinfo {year} {2022})},\ \Eprint {http://arxiv.org/abs/2112.15214} {arXiv:2112.15214 [hep-ph]} \BibitemShut {NoStop}%
\bibitem [{\citenamefont {Schwartz}(2014)}]{Schwartz:2014sze}%
  \BibitemOpen
  \bibfield  {author} {\bibinfo {author} {\bibfnamefont {M.~D.}\ \bibnamefont {Schwartz}},\ }\href@noop {} {\emph {\bibinfo {title} {{Quantum Field Theory and the Standard Model}}}}\ (\bibinfo  {publisher} {Cambridge University Press},\ \bibinfo {year} {2014})\BibitemShut {NoStop}%
\bibitem [{\citenamefont {Greene}\ and\ \citenamefont {Kofman}(1999)}]{Greene:1998nh}%
  \BibitemOpen
  \bibfield  {author} {\bibinfo {author} {\bibfnamefont {P.~B.}\ \bibnamefont {Greene}}\ and\ \bibinfo {author} {\bibfnamefont {L.}~\bibnamefont {Kofman}},\ }\href {\doibase 10.1016/S0370-2693(99)00020-9} {\bibfield  {journal} {\bibinfo  {journal} {Phys. Lett. B}\ }\textbf {\bibinfo {volume} {448}},\ \bibinfo {pages} {6} (\bibinfo {year} {1999})},\ \Eprint {http://arxiv.org/abs/hep-ph/9807339} {arXiv:hep-ph/9807339} \BibitemShut {NoStop}%
\bibitem [{\citenamefont {Adshead}\ \emph {et~al.}(2018)\citenamefont {Adshead}, \citenamefont {Pearce}, \citenamefont {Peloso}, \citenamefont {Roberts},\ and\ \citenamefont {Sorbo}}]{Adshead:2018oaa}%
  \BibitemOpen
  \bibfield  {author} {\bibinfo {author} {\bibfnamefont {P.}~\bibnamefont {Adshead}}, \bibinfo {author} {\bibfnamefont {L.}~\bibnamefont {Pearce}}, \bibinfo {author} {\bibfnamefont {M.}~\bibnamefont {Peloso}}, \bibinfo {author} {\bibfnamefont {M.~A.}\ \bibnamefont {Roberts}}, \ and\ \bibinfo {author} {\bibfnamefont {L.}~\bibnamefont {Sorbo}},\ }\href {\doibase 10.1088/1475-7516/2018/06/020} {\bibfield  {journal} {\bibinfo  {journal} {JCAP}\ }\textbf {\bibinfo {volume} {06}},\ \bibinfo {pages} {020} (\bibinfo {year} {2018})},\ \Eprint {http://arxiv.org/abs/1803.04501} {arXiv:1803.04501 [astro-ph.CO]} \BibitemShut {NoStop}%
\bibitem [{\citenamefont {Maleknejad}(2020{\natexlab{a}})}]{Maleknejad:2019hdr}%
  \BibitemOpen
  \bibfield  {author} {\bibinfo {author} {\bibfnamefont {A.}~\bibnamefont {Maleknejad}},\ }\href {\doibase 10.1007/JHEP07(2020)154} {\bibfield  {journal} {\bibinfo  {journal} {JHEP}\ }\textbf {\bibinfo {volume} {07}},\ \bibinfo {pages} {154} (\bibinfo {year} {2020}{\natexlab{a}})},\ \Eprint {http://arxiv.org/abs/1909.11545} {arXiv:1909.11545 [hep-th]} \BibitemShut {NoStop}%
\bibitem [{\citenamefont {Maleknejad}(2021)}]{Maleknejad:2020yys}%
  \BibitemOpen
  \bibfield  {author} {\bibinfo {author} {\bibfnamefont {A.}~\bibnamefont {Maleknejad}},\ }\href {\doibase 10.1103/PhysRevD.104.083518} {\bibfield  {journal} {\bibinfo  {journal} {Phys. Rev. D}\ }\textbf {\bibinfo {volume} {104}},\ \bibinfo {pages} {083518} (\bibinfo {year} {2021})},\ \Eprint {http://arxiv.org/abs/2012.11516} {arXiv:2012.11516 [hep-ph]} \BibitemShut {NoStop}%
\bibitem [{\citenamefont {Maleknejad}(2020{\natexlab{b}})}]{Maleknejad:2020pec}%
  \BibitemOpen
  \bibfield  {author} {\bibinfo {author} {\bibfnamefont {A.}~\bibnamefont {Maleknejad}},\ }\href {\doibase 10.1007/JHEP06(2021)113} {\bibfield  {journal} {\bibinfo  {journal} {JHEP}\ }\textbf {\bibinfo {volume} {21}},\ \bibinfo {pages} {113} (\bibinfo {year} {2020}{\natexlab{b}})},\ \Eprint {http://arxiv.org/abs/2103.14611} {arXiv:2103.14611 [hep-ph]} \BibitemShut {NoStop}%
\bibitem [{\citenamefont {Alexander}\ \emph {et~al.}(2006)\citenamefont {Alexander}, \citenamefont {Peskin},\ and\ \citenamefont {Sheikh-Jabbari}}]{Alexander:2004us}%
  \BibitemOpen
  \bibfield  {author} {\bibinfo {author} {\bibfnamefont {S.~H.-S.}\ \bibnamefont {Alexander}}, \bibinfo {author} {\bibfnamefont {M.~E.}\ \bibnamefont {Peskin}}, \ and\ \bibinfo {author} {\bibfnamefont {M.~M.}\ \bibnamefont {Sheikh-Jabbari}},\ }\href {\doibase 10.1103/PhysRevLett.96.081301} {\bibfield  {journal} {\bibinfo  {journal} {Phys. Rev. Lett.}\ }\textbf {\bibinfo {volume} {96}},\ \bibinfo {pages} {081301} (\bibinfo {year} {2006})},\ \Eprint {http://arxiv.org/abs/hep-th/0403069} {arXiv:hep-th/0403069} \BibitemShut {NoStop}%
\bibitem [{\citenamefont {Maleknejad}(2016{\natexlab{a}})}]{Maleknejad:2016dci}%
  \BibitemOpen
  \bibfield  {author} {\bibinfo {author} {\bibfnamefont {A.}~\bibnamefont {Maleknejad}},\ }\href {\doibase 10.1088/1475-7516/2016/12/027} {\bibfield  {journal} {\bibinfo  {journal} {JCAP}\ }\textbf {\bibinfo {volume} {12}},\ \bibinfo {pages} {027} (\bibinfo {year} {2016}{\natexlab{a}})},\ \Eprint {http://arxiv.org/abs/1604.06520} {arXiv:1604.06520 [hep-ph]} \BibitemShut {NoStop}%
\bibitem [{\citenamefont {Maleknejad}(2014)}]{Maleknejad:2014wsa}%
  \BibitemOpen
  \bibfield  {author} {\bibinfo {author} {\bibfnamefont {A.}~\bibnamefont {Maleknejad}},\ }\href {\doibase 10.1103/PhysRevD.90.023542} {\bibfield  {journal} {\bibinfo  {journal} {Phys. Rev. D}\ }\textbf {\bibinfo {volume} {90}},\ \bibinfo {pages} {023542} (\bibinfo {year} {2014})},\ \Eprint {http://arxiv.org/abs/1401.7628} {arXiv:1401.7628 [hep-th]} \BibitemShut {NoStop}%
\bibitem [{\citenamefont {Adshead}\ and\ \citenamefont {Sfakianakis}(2016)}]{Adshead:2015jza}%
  \BibitemOpen
  \bibfield  {author} {\bibinfo {author} {\bibfnamefont {P.}~\bibnamefont {Adshead}}\ and\ \bibinfo {author} {\bibfnamefont {E.~I.}\ \bibnamefont {Sfakianakis}},\ }\href {\doibase 10.1103/PhysRevLett.116.091301} {\bibfield  {journal} {\bibinfo  {journal} {Phys. Rev. Lett.}\ }\textbf {\bibinfo {volume} {116}},\ \bibinfo {pages} {091301} (\bibinfo {year} {2016})},\ \Eprint {http://arxiv.org/abs/1508.00881} {arXiv:1508.00881 [hep-ph]} \BibitemShut {NoStop}%
\bibitem [{\citenamefont {Caldwell}\ and\ \citenamefont {Devulder}(2018)}]{Caldwell:2017chz}%
  \BibitemOpen
  \bibfield  {author} {\bibinfo {author} {\bibfnamefont {R.~R.}\ \bibnamefont {Caldwell}}\ and\ \bibinfo {author} {\bibfnamefont {C.}~\bibnamefont {Devulder}},\ }\href {\doibase 10.1103/PhysRevD.97.023532} {\bibfield  {journal} {\bibinfo  {journal} {Phys. Rev. D}\ }\textbf {\bibinfo {volume} {97}},\ \bibinfo {pages} {023532} (\bibinfo {year} {2018})},\ \Eprint {http://arxiv.org/abs/1706.03765} {arXiv:1706.03765 [astro-ph.CO]} \BibitemShut {NoStop}%
\bibitem [{\citenamefont {Campos}\ and\ \citenamefont {Verdaguer}(1992)}]{Campos:1991ff}%
  \BibitemOpen
  \bibfield  {author} {\bibinfo {author} {\bibfnamefont {A.}~\bibnamefont {Campos}}\ and\ \bibinfo {author} {\bibfnamefont {E.}~\bibnamefont {Verdaguer}},\ }\href {\doibase 10.1103/PhysRevD.45.4428} {\bibfield  {journal} {\bibinfo  {journal} {Phys. Rev. D}\ }\textbf {\bibinfo {volume} {45}},\ \bibinfo {pages} {4428} (\bibinfo {year} {1992})}\BibitemShut {NoStop}%
\bibitem [{\citenamefont {Sorbo}(2011)}]{Sorbo:2011rz}%
  \BibitemOpen
  \bibfield  {author} {\bibinfo {author} {\bibfnamefont {L.}~\bibnamefont {Sorbo}},\ }\href {\doibase 10.1088/1475-7516/2011/06/003} {\bibfield  {journal} {\bibinfo  {journal} {JCAP}\ }\textbf {\bibinfo {volume} {06}},\ \bibinfo {pages} {003} (\bibinfo {year} {2011})},\ \Eprint {http://arxiv.org/abs/1101.1525} {arXiv:1101.1525 [astro-ph.CO]} \BibitemShut {NoStop}%
\bibitem [{\citenamefont {Maleknejad}(2016{\natexlab{b}})}]{Maleknejad:2016qjz}%
  \BibitemOpen
  \bibfield  {author} {\bibinfo {author} {\bibfnamefont {A.}~\bibnamefont {Maleknejad}},\ }\href {\doibase 10.1007/JHEP07(2016)104} {\bibfield  {journal} {\bibinfo  {journal} {JHEP}\ }\textbf {\bibinfo {volume} {07}},\ \bibinfo {pages} {104} (\bibinfo {year} {2016}{\natexlab{b}})},\ \Eprint {http://arxiv.org/abs/1604.03327} {arXiv:1604.03327 [hep-ph]} \BibitemShut {NoStop}%
\bibitem [{\citenamefont {Komatsu}(2022)}]{Komatsu:2022nvu}%
  \BibitemOpen
  \bibfield  {author} {\bibinfo {author} {\bibfnamefont {E.}~\bibnamefont {Komatsu}},\ }\href {\doibase 10.1038/s42254-022-00452-4} {\bibfield  {journal} {\bibinfo  {journal} {Nature Rev. Phys.}\ }\textbf {\bibinfo {volume} {4}},\ \bibinfo {pages} {452} (\bibinfo {year} {2022})},\ \Eprint {http://arxiv.org/abs/2202.13919} {arXiv:2202.13919 [astro-ph.CO]} \BibitemShut {NoStop}%
\bibitem [{\citenamefont {Witten}(1984)}]{Witten:1984rs}%
  \BibitemOpen
  \bibfield  {author} {\bibinfo {author} {\bibfnamefont {E.}~\bibnamefont {Witten}},\ }\href {\doibase 10.1103/PhysRevD.30.272} {\bibfield  {journal} {\bibinfo  {journal} {Phys. Rev. D}\ }\textbf {\bibinfo {volume} {30}},\ \bibinfo {pages} {272} (\bibinfo {year} {1984})}\BibitemShut {NoStop}%
\bibitem [{\citenamefont {Schwaller}(2015)}]{Schwaller:2015tja}%
  \BibitemOpen
  \bibfield  {author} {\bibinfo {author} {\bibfnamefont {P.}~\bibnamefont {Schwaller}},\ }\href {\doibase 10.1103/PhysRevLett.115.181101} {\bibfield  {journal} {\bibinfo  {journal} {Phys. Rev. Lett.}\ }\textbf {\bibinfo {volume} {115}},\ \bibinfo {pages} {181101} (\bibinfo {year} {2015})},\ \Eprint {http://arxiv.org/abs/1504.07263} {arXiv:1504.07263 [hep-ph]} \BibitemShut {NoStop}%
\bibitem [{\citenamefont {Caprini}\ \emph {et~al.}(2016)\citenamefont {Caprini} \emph {et~al.}}]{Caprini:2015zlo}%
  \BibitemOpen
  \bibfield  {author} {\bibinfo {author} {\bibfnamefont {C.}~\bibnamefont {Caprini}} \emph {et~al.},\ }\href {\doibase 10.1088/1475-7516/2016/04/001} {\bibfield  {journal} {\bibinfo  {journal} {JCAP}\ }\textbf {\bibinfo {volume} {04}},\ \bibinfo {pages} {001} (\bibinfo {year} {2016})},\ \Eprint {http://arxiv.org/abs/1512.06239} {arXiv:1512.06239 [astro-ph.CO]} \BibitemShut {NoStop}%
\bibitem [{\citenamefont {Roper~Pol}\ \emph {et~al.}(2023)\citenamefont {Roper~Pol}, \citenamefont {Neronov}, \citenamefont {Caprini}, \citenamefont {Boyer},\ and\ \citenamefont {Semikoz}}]{RoperPol:2023bqa}%
  \BibitemOpen
  \bibfield  {author} {\bibinfo {author} {\bibfnamefont {A.}~\bibnamefont {Roper~Pol}}, \bibinfo {author} {\bibfnamefont {A.}~\bibnamefont {Neronov}}, \bibinfo {author} {\bibfnamefont {C.}~\bibnamefont {Caprini}}, \bibinfo {author} {\bibfnamefont {T.}~\bibnamefont {Boyer}}, \ and\ \bibinfo {author} {\bibfnamefont {D.}~\bibnamefont {Semikoz}},\ }\href@noop {} {\  (\bibinfo {year} {2023})},\ \Eprint {http://arxiv.org/abs/2307.10744} {arXiv:2307.10744 [astro-ph.CO]} \BibitemShut {NoStop}%
\bibitem [{\citenamefont {Brandenburg}\ \emph {et~al.}(2021)\citenamefont {Brandenburg}, \citenamefont {He}, \citenamefont {Kahniashvili}, \citenamefont {Rheinhardt},\ and\ \citenamefont {Schober}}]{Brandenburg:2021aln}%
  \BibitemOpen
  \bibfield  {author} {\bibinfo {author} {\bibfnamefont {A.}~\bibnamefont {Brandenburg}}, \bibinfo {author} {\bibfnamefont {Y.}~\bibnamefont {He}}, \bibinfo {author} {\bibfnamefont {T.}~\bibnamefont {Kahniashvili}}, \bibinfo {author} {\bibfnamefont {M.}~\bibnamefont {Rheinhardt}}, \ and\ \bibinfo {author} {\bibfnamefont {J.}~\bibnamefont {Schober}},\ }\href {\doibase 10.3847/1538-4357/abe4d7} {\bibfield  {journal} {\bibinfo  {journal} {Astrophys. J.}\ }\textbf {\bibinfo {volume} {911}},\ \bibinfo {pages} {110} (\bibinfo {year} {2021})},\ \Eprint {http://arxiv.org/abs/2101.08178} {arXiv:2101.08178 [astro-ph.CO]} \BibitemShut {NoStop}%
\bibitem [{\citenamefont {Roper~Pol}\ \emph {et~al.}(2022{\natexlab{a}})\citenamefont {Roper~Pol}, \citenamefont {Mandal}, \citenamefont {Brandenburg},\ and\ \citenamefont {Kahniashvili}}]{RoperPol:2021xnd}%
  \BibitemOpen
  \bibfield  {author} {\bibinfo {author} {\bibfnamefont {A.}~\bibnamefont {Roper~Pol}}, \bibinfo {author} {\bibfnamefont {S.}~\bibnamefont {Mandal}}, \bibinfo {author} {\bibfnamefont {A.}~\bibnamefont {Brandenburg}}, \ and\ \bibinfo {author} {\bibfnamefont {T.}~\bibnamefont {Kahniashvili}},\ }\href {\doibase 10.1088/1475-7516/2022/04/019} {\bibfield  {journal} {\bibinfo  {journal} {JCAP}\ }\textbf {\bibinfo {volume} {04}},\ \bibinfo {pages} {019} (\bibinfo {year} {2022}{\natexlab{a}})},\ \Eprint {http://arxiv.org/abs/2107.05356} {arXiv:2107.05356 [gr-qc]} \BibitemShut {NoStop}%
\bibitem [{\citenamefont {Adshead}\ \emph {et~al.}(2020)\citenamefont {Adshead}, \citenamefont {Giblin}, \citenamefont {Pieroni},\ and\ \citenamefont {Weiner}}]{Adshead:2019igv}%
  \BibitemOpen
  \bibfield  {author} {\bibinfo {author} {\bibfnamefont {P.}~\bibnamefont {Adshead}}, \bibinfo {author} {\bibfnamefont {J.~T.}\ \bibnamefont {Giblin}}, \bibinfo {author} {\bibfnamefont {M.}~\bibnamefont {Pieroni}}, \ and\ \bibinfo {author} {\bibfnamefont {Z.~J.}\ \bibnamefont {Weiner}},\ }\href {\doibase 10.1103/PhysRevLett.124.171301} {\bibfield  {journal} {\bibinfo  {journal} {Phys. Rev. Lett.}\ }\textbf {\bibinfo {volume} {124}},\ \bibinfo {pages} {171301} (\bibinfo {year} {2020})},\ \Eprint {http://arxiv.org/abs/1909.12843} {arXiv:1909.12843 [astro-ph.CO]} \BibitemShut {NoStop}%
\bibitem [{\citenamefont {Figueroa}\ \emph {et~al.}(2022)\citenamefont {Figueroa}, \citenamefont {Florio}, \citenamefont {Loayza},\ and\ \citenamefont {Pieroni}}]{Figueroa:2022iho}%
  \BibitemOpen
  \bibfield  {author} {\bibinfo {author} {\bibfnamefont {D.~G.}\ \bibnamefont {Figueroa}}, \bibinfo {author} {\bibfnamefont {A.}~\bibnamefont {Florio}}, \bibinfo {author} {\bibfnamefont {N.}~\bibnamefont {Loayza}}, \ and\ \bibinfo {author} {\bibfnamefont {M.}~\bibnamefont {Pieroni}},\ }\href {\doibase 10.1103/PhysRevD.106.063522} {\bibfield  {journal} {\bibinfo  {journal} {Phys. Rev. D}\ }\textbf {\bibinfo {volume} {106}},\ \bibinfo {pages} {063522} (\bibinfo {year} {2022})},\ \Eprint {http://arxiv.org/abs/2202.05805} {arXiv:2202.05805 [astro-ph.CO]} \BibitemShut {NoStop}%
\bibitem [{\citenamefont {Hindmarsh}\ and\ \citenamefont {Kibble}(1995)}]{Hindmarsh:1994re}%
  \BibitemOpen
  \bibfield  {author} {\bibinfo {author} {\bibfnamefont {M.~B.}\ \bibnamefont {Hindmarsh}}\ and\ \bibinfo {author} {\bibfnamefont {T.~W.~B.}\ \bibnamefont {Kibble}},\ }\href {\doibase 10.1088/0034-4885/58/5/001} {\bibfield  {journal} {\bibinfo  {journal} {Rept. Prog. Phys.}\ }\textbf {\bibinfo {volume} {58}},\ \bibinfo {pages} {477} (\bibinfo {year} {1995})},\ \Eprint {http://arxiv.org/abs/hep-ph/9411342} {arXiv:hep-ph/9411342} \BibitemShut {NoStop}%
\bibitem [{\citenamefont {Auclair}\ \emph {et~al.}(2020)\citenamefont {Auclair} \emph {et~al.}}]{Auclair:2019wcv}%
  \BibitemOpen
  \bibfield  {author} {\bibinfo {author} {\bibfnamefont {P.}~\bibnamefont {Auclair}} \emph {et~al.},\ }\href {\doibase 10.1088/1475-7516/2020/04/034} {\bibfield  {journal} {\bibinfo  {journal} {JCAP}\ }\textbf {\bibinfo {volume} {04}},\ \bibinfo {pages} {034} (\bibinfo {year} {2020})},\ \Eprint {http://arxiv.org/abs/1909.00819} {arXiv:1909.00819 [astro-ph.CO]} \BibitemShut {NoStop}%
\bibitem [{\citenamefont {Caprini}\ \emph {et~al.}(2009{\natexlab{a}})\citenamefont {Caprini}, \citenamefont {Durrer},\ and\ \citenamefont {Servant}}]{Caprini:2009yp}%
  \BibitemOpen
  \bibfield  {author} {\bibinfo {author} {\bibfnamefont {C.}~\bibnamefont {Caprini}}, \bibinfo {author} {\bibfnamefont {R.}~\bibnamefont {Durrer}}, \ and\ \bibinfo {author} {\bibfnamefont {G.}~\bibnamefont {Servant}},\ }\href {\doibase 10.1088/1475-7516/2009/12/024} {\bibfield  {journal} {\bibinfo  {journal} {JCAP}\ }\textbf {\bibinfo {volume} {12}},\ \bibinfo {pages} {024} (\bibinfo {year} {2009}{\natexlab{a}})},\ \Eprint {http://arxiv.org/abs/0909.0622} {arXiv:0909.0622 [astro-ph.CO]} \BibitemShut {NoStop}%
\bibitem [{\citenamefont {Caprini}\ and\ \citenamefont {Figueroa}(2018)}]{Caprini:2018mtu}%
  \BibitemOpen
  \bibfield  {author} {\bibinfo {author} {\bibfnamefont {C.}~\bibnamefont {Caprini}}\ and\ \bibinfo {author} {\bibfnamefont {D.~G.}\ \bibnamefont {Figueroa}},\ }\href {\doibase 10.1088/1361-6382/aac608} {\bibfield  {journal} {\bibinfo  {journal} {Class. Quant. Grav.}\ }\textbf {\bibinfo {volume} {35}},\ \bibinfo {pages} {163001} (\bibinfo {year} {2018})},\ \Eprint {http://arxiv.org/abs/1801.04268} {arXiv:1801.04268 [astro-ph.CO]} \BibitemShut {NoStop}%
\bibitem [{\citenamefont {Maleknejad}\ and\ \citenamefont {Kopp}(2025)}]{JHEP}%
  \BibitemOpen
  \bibfield  {author} {\bibinfo {author} {\bibfnamefont {A.}~\bibnamefont {Maleknejad}}\ and\ \bibinfo {author} {\bibfnamefont {J.}~\bibnamefont {Kopp}},\ }\href {\doibase 10.1007/JHEP01(2025)023} {\bibfield  {journal} {\bibinfo  {journal} {JHEP}\ }\textbf {\bibinfo {volume} {01}},\ \bibinfo {pages} {023} (\bibinfo {year} {2025})},\ \Eprint {http://arxiv.org/abs/2406.01534} {arXiv:2406.01534 [hep-th]} \BibitemShut {NoStop}%
\bibitem [{Note1()}]{Note1}%
  \BibitemOpen
  \bibinfo {note} {Working with circular GW polarization states is more convenient here than using the plus and cross polarizations as a basis, even though the latter is more common in the GW literature.}\BibitemShut {Stop}%
\bibitem [{Note2()}]{Note2}%
  \BibitemOpen
  \bibinfo {note} {Note that at second order in gravitational wave perturbations, the perturbed metric is $\delta g_{ij}=a^2(h_{ij} + \protect \frac 12 h_{ik} h_{jk})$.}\BibitemShut {Stop}%
\bibitem [{Note3()}]{Note3}%
  \BibitemOpen
  \bibinfo {note} {Given that helicity and chirality are equivalent for free fermions, the zeroth-order solution for left-handed Weyl fermions is similar, but with ${\protect \bf {E}}^+_{{\protect \bf {k}}}$ replaced by ${\protect \bf {E}}^-_{{\protect \bf {k}}}$.}\BibitemShut {Stop}%
\bibitem [{\citenamefont {Alvarez-Gaume}\ and\ \citenamefont {Witten}(1984)}]{Alvarez-Gaume:1983ihn}%
  \BibitemOpen
  \bibfield  {author} {\bibinfo {author} {\bibfnamefont {L.}~\bibnamefont {Alvarez-Gaume}}\ and\ \bibinfo {author} {\bibfnamefont {E.}~\bibnamefont {Witten}},\ }\href {\doibase 10.1016/0550-3213(84)90066-X} {\bibfield  {journal} {\bibinfo  {journal} {Nucl. Phys. B}\ }\textbf {\bibinfo {volume} {234}},\ \bibinfo {pages} {269} (\bibinfo {year} {1984})}\BibitemShut {NoStop}%
\bibitem [{\citenamefont {Maleknejad}(2024)}]{Maleknejad:2024vvf}%
  \BibitemOpen
  \bibfield  {author} {\bibinfo {author} {\bibfnamefont {A.}~\bibnamefont {Maleknejad}},\ }\href@noop {} {\  (\bibinfo {year} {2024})},\ \Eprint {http://arxiv.org/abs/2412.09490} {arXiv:2412.09490 [hep-ph]} \BibitemShut {NoStop}%
\bibitem [{Note4()}]{Note4}%
  \BibitemOpen
  \bibinfo {note} {We have already performed the quantum expectation value over the fermionic fields, and the remaining averaging on GW is purely stochastic.}\BibitemShut {Stop}%
\bibitem [{Note5()}]{Note5}%
  \BibitemOpen
  \bibinfo {note} {A long-wavelength GW mode (homogeneous tensor mode) $h^\protect \text {long}_{ij}(t,{\protect \bf {x}}) \approx C_{ij}(t)$, can be recast as a change in coordinates $h^{\protect \text {long}}_{ij} \DOTSB \mapstochar \rightarrow 0$ and $x^i \DOTSB \mapstochar \rightarrow x^{i'} = e^{h^{\protect \text {long}}_{ij}} x^j$ \cite {Weinberg:2008zzc,Senatore:2012nq}. In other words, it can be considered as a large gauge transformation. As a result, after this coordinate transformation, a short-wavelength fermion mode in the presence of long-wavelength GWs becomes a free fermion, i.e., \protect \smash {$\Psi _\protect \text {short}(t,{\protect \bf {x}}) \protect \big \rvert _{h_{ij}^{\protect \text {long}}} = \Psi '(t,{\protect \bf {x}}') \protect \big \rvert _0$}, which implies $\langle \rho _{\psi ,\protect \text {short}}(x) \rangle \protect \big \rvert _{h_{ij}^\protect \text {long}} = \langle \rho _{\psi ,\protect \text {short}}(x') \rangle \protect \big \rvert _0 = 0$.}\BibitemShut {Stop}%
\bibitem [{Note6()}]{Note6}%
  \BibitemOpen
  \bibinfo {note} {The energy density of GWs is defined as $\rho _\protect \text {gw} = \protect \frac {1}{32\pi G} \protect \big \langle \protect \dot {h}_{ij}\protect \, \protect \dot {h}^{ij}\protect \big \rangle $ and its spectral energy density is $\Omega _\protect \text {gw} = \protect \frac {1}{M_{\protect \mbox {\relax \protect \fontsize {5}{6}\protect \selectfont {Pl}}}^2} \protect \frac {d\rho _\protect \text {gw}}{3H^2 d\ln q} $.}\BibitemShut {Stop}%
\bibitem [{\citenamefont {Durrer}(2010)}]{Durrer:2010xc}%
  \BibitemOpen
  \bibfield  {author} {\bibinfo {author} {\bibfnamefont {R.}~\bibnamefont {Durrer}},\ }\href {\doibase 10.1088/1742-6596/222/1/012021} {\bibfield  {journal} {\bibinfo  {journal} {J. Phys. Conf. Ser.}\ }\textbf {\bibinfo {volume} {222}},\ \bibinfo {pages} {012021} (\bibinfo {year} {2010})},\ \Eprint {http://arxiv.org/abs/1002.1389} {arXiv:1002.1389 [astro-ph.CO]} \BibitemShut {NoStop}%
\bibitem [{\citenamefont {Roper~Pol}\ \emph {et~al.}(2022{\natexlab{b}})\citenamefont {Roper~Pol}, \citenamefont {Caprini}, \citenamefont {Neronov},\ and\ \citenamefont {Semikoz}}]{RoperPol:2022iel}%
  \BibitemOpen
  \bibfield  {author} {\bibinfo {author} {\bibfnamefont {A.}~\bibnamefont {Roper~Pol}}, \bibinfo {author} {\bibfnamefont {C.}~\bibnamefont {Caprini}}, \bibinfo {author} {\bibfnamefont {A.}~\bibnamefont {Neronov}}, \ and\ \bibinfo {author} {\bibfnamefont {D.}~\bibnamefont {Semikoz}},\ }\href {\doibase 10.1103/PhysRevD.105.123502} {\bibfield  {journal} {\bibinfo  {journal} {Phys. Rev. D}\ }\textbf {\bibinfo {volume} {105}},\ \bibinfo {pages} {123502} (\bibinfo {year} {2022}{\natexlab{b}})},\ \Eprint {http://arxiv.org/abs/2201.05630} {arXiv:2201.05630 [astro-ph.CO]} \BibitemShut {NoStop}%
\bibitem [{\citenamefont {Breitbach}\ \emph {et~al.}(2019)\citenamefont {Breitbach}, \citenamefont {Kopp}, \citenamefont {Madge}, \citenamefont {Opferkuch},\ and\ \citenamefont {Schwaller}}]{Breitbach:2018ddu}%
  \BibitemOpen
  \bibfield  {author} {\bibinfo {author} {\bibfnamefont {M.}~\bibnamefont {Breitbach}}, \bibinfo {author} {\bibfnamefont {J.}~\bibnamefont {Kopp}}, \bibinfo {author} {\bibfnamefont {E.}~\bibnamefont {Madge}}, \bibinfo {author} {\bibfnamefont {T.}~\bibnamefont {Opferkuch}}, \ and\ \bibinfo {author} {\bibfnamefont {P.}~\bibnamefont {Schwaller}},\ }\href {\doibase 10.1088/1475-7516/2019/07/007} {\bibfield  {journal} {\bibinfo  {journal} {JCAP}\ }\textbf {\bibinfo {volume} {07}},\ \bibinfo {pages} {007} (\bibinfo {year} {2019})},\ \Eprint {http://arxiv.org/abs/1811.11175} {arXiv:1811.11175 [hep-ph]} \BibitemShut {NoStop}%
\bibitem [{\citenamefont {Breitbach}(2018)}]{Breitbach:2018kma}%
  \BibitemOpen
  \bibfield  {author} {\bibinfo {author} {\bibfnamefont {M.}~\bibnamefont {Breitbach}},\ }\emph {\bibinfo {title} {{Gravitational Waves from Cosmological Phase Transitions}}},\ \href@noop {} {Master's thesis},\ \bibinfo  {school} {Mainz U.} (\bibinfo {year} {2018}),\ \Eprint {http://arxiv.org/abs/2204.09661} {arXiv:2204.09661 [astro-ph.CO]} \BibitemShut {NoStop}%
\bibitem [{\citenamefont {Scully}\ and\ \citenamefont {Zubairy}(1997)}]{Scully_Zubairy_1997}%
  \BibitemOpen
  \bibfield  {author} {\bibinfo {author} {\bibfnamefont {M.~O.}\ \bibnamefont {Scully}}\ and\ \bibinfo {author} {\bibfnamefont {M.~S.}\ \bibnamefont {Zubairy}},\ }\href@noop {} {\emph {\bibinfo {title} {{Quantum Optics}}}}\ (\bibinfo {year} {1997})\BibitemShut {NoStop}%
\bibitem [{\citenamefont {Caprini}\ \emph {et~al.}(2009{\natexlab{b}})\citenamefont {Caprini}, \citenamefont {Durrer}, \citenamefont {Konstandin},\ and\ \citenamefont {Servant}}]{Caprini:2009fx}%
  \BibitemOpen
  \bibfield  {author} {\bibinfo {author} {\bibfnamefont {C.}~\bibnamefont {Caprini}}, \bibinfo {author} {\bibfnamefont {R.}~\bibnamefont {Durrer}}, \bibinfo {author} {\bibfnamefont {T.}~\bibnamefont {Konstandin}}, \ and\ \bibinfo {author} {\bibfnamefont {G.}~\bibnamefont {Servant}},\ }\href {\doibase 10.1103/PhysRevD.79.083519} {\bibfield  {journal} {\bibinfo  {journal} {Phys. Rev. D}\ }\textbf {\bibinfo {volume} {79}},\ \bibinfo {pages} {083519} (\bibinfo {year} {2009}{\natexlab{b}})},\ \Eprint {http://arxiv.org/abs/0901.1661} {arXiv:0901.1661 [astro-ph.CO]} \BibitemShut {NoStop}%
\bibitem [{Note7()}]{Note7}%
  \BibitemOpen
  \bibinfo {note} {Note that $\Omega _\protect \text {peak}$ may be related to $T_*$ and the GW background properties in a model-dependent manner. This work focuses on a broadly applicable framework, leaving a detailed exploration to future numerical studies.}\BibitemShut {Stop}%
\bibitem [{\citenamefont {Maggiore}(2018)}]{Maggiore:2018sht}%
  \BibitemOpen
  \bibfield  {author} {\bibinfo {author} {\bibfnamefont {M.}~\bibnamefont {Maggiore}},\ }\href@noop {} {\emph {\bibinfo {title} {{Gravitational Waves. Vol. 2: Astrophysics and Cosmology}}}}\ (\bibinfo  {publisher} {Oxford University Press},\ \bibinfo {year} {2018})\BibitemShut {NoStop}%
\bibitem [{\citenamefont {Maggiore}\ \emph {et~al.}(2020)\citenamefont {Maggiore} \emph {et~al.}}]{Maggiore:2019uih}%
  \BibitemOpen
  \bibfield  {author} {\bibinfo {author} {\bibfnamefont {M.}~\bibnamefont {Maggiore}} \emph {et~al.},\ }\href {\doibase 10.1088/1475-7516/2020/03/050} {\bibfield  {journal} {\bibinfo  {journal} {JCAP}\ }\textbf {\bibinfo {volume} {03}},\ \bibinfo {pages} {050} (\bibinfo {year} {2020})},\ \Eprint {http://arxiv.org/abs/1912.02622} {arXiv:1912.02622 [astro-ph.CO]} \BibitemShut {NoStop}%
\bibitem [{\citenamefont {Evans}\ \emph {et~al.}(2021)\citenamefont {Evans} \emph {et~al.}}]{Evans:2021gyd}%
  \BibitemOpen
  \bibfield  {author} {\bibinfo {author} {\bibfnamefont {M.}~\bibnamefont {Evans}} \emph {et~al.},\ }\href@noop {} {\  (\bibinfo {year} {2021})},\ \Eprint {http://arxiv.org/abs/2109.09882} {arXiv:2109.09882 [astro-ph.IM]} \BibitemShut {NoStop}%
\bibitem [{\citenamefont {Aggarwal}\ \emph {et~al.}(2021)\citenamefont {Aggarwal} \emph {et~al.}}]{Aggarwal:2020olq}%
  \BibitemOpen
  \bibfield  {author} {\bibinfo {author} {\bibfnamefont {N.}~\bibnamefont {Aggarwal}} \emph {et~al.},\ }\href {\doibase 10.1007/s41114-021-00032-5} {\bibfield  {journal} {\bibinfo  {journal} {Living Rev. Rel.}\ }\textbf {\bibinfo {volume} {24}},\ \bibinfo {pages} {4} (\bibinfo {year} {2021})},\ \Eprint {http://arxiv.org/abs/2011.12414} {arXiv:2011.12414 [gr-qc]} \BibitemShut {NoStop}%
\bibitem [{\citenamefont {Maggiore}(2007)}]{Maggiore:2007ulw}%
  \BibitemOpen
  \bibfield  {author} {\bibinfo {author} {\bibfnamefont {M.}~\bibnamefont {Maggiore}},\ }\href {\doibase 10.1093/acprof:oso/9780198570745.001.0001} {\emph {\bibinfo {title} {{Gravitational Waves. Vol. 1: Theory and Experiments}}}}\ (\bibinfo  {publisher} {Oxford University Press},\ \bibinfo {year} {2007})\BibitemShut {NoStop}%
\bibitem [{\citenamefont {Strominger}(2017)}]{Strominger:2017zoo}%
  \BibitemOpen
  \bibfield  {author} {\bibinfo {author} {\bibfnamefont {A.}~\bibnamefont {Strominger}},\ }\href@noop {} {\emph {\bibinfo {title} {{Lectures on the Infrared Structure of Gravity and Gauge Theory}}}}\ (\bibinfo {year} {2017})\ \Eprint {http://arxiv.org/abs/1703.05448} {arXiv:1703.05448 [hep-th]} \BibitemShut {NoStop}%
\bibitem [{\citenamefont {Weinberg}(2008)}]{Weinberg:2008zzc}%
  \BibitemOpen
  \bibfield  {author} {\bibinfo {author} {\bibfnamefont {S.}~\bibnamefont {Weinberg}},\ }\href@noop {} {\emph {\bibinfo {title} {{Cosmology}}}}\ (\bibinfo {year} {2008})\BibitemShut {NoStop}%
\bibitem [{\citenamefont {Senatore}\ and\ \citenamefont {Zaldarriaga}(2013)}]{Senatore:2012nq}%
  \BibitemOpen
  \bibfield  {author} {\bibinfo {author} {\bibfnamefont {L.}~\bibnamefont {Senatore}}\ and\ \bibinfo {author} {\bibfnamefont {M.}~\bibnamefont {Zaldarriaga}},\ }\href {\doibase 10.1007/JHEP01(2013)109} {\bibfield  {journal} {\bibinfo  {journal} {JHEP}\ }\textbf {\bibinfo {volume} {01}},\ \bibinfo {pages} {109} (\bibinfo {year} {2013})},\ \Eprint {http://arxiv.org/abs/1203.6354} {arXiv:1203.6354 [hep-th]} \BibitemShut {NoStop}%
\end{thebibliography}%


\end{document}